\documentclass[]{aa}
\usepackage{txfonts}
\usepackage{graphicx}
\usepackage{float}
\usepackage{lscape}
\usepackage{natbib}
\usepackage{url}
\usepackage{array}
\bibpunct{(}{)}{;}{a}{}{,}

\newcommand{\nf}{\mbox{\texttt{Nightfall}}}

\newcommand{\Teff}{\mbox{$T_{\rm eff}$}}
\newcommand{\Lg}{\mbox{$\log\,g$}}

\newcommand{\Msol}{\mbox{$M_{\odot}$}}
\newcommand{\Rsol}{\mbox{$R_{\odot}$}}

\begin{document}

\title{Investigation of transit-selected exoplanet candidates from the
  MACHO survey\thanks{Based on observations made with ESO Telescopes
  at the La Silla or Paranal Observatories under programme ID
  075.C-0526(A)}}
  \titlerunning{Investigation of transit-selected exoplanet candidates} 
  \author{S.~D. H\"ugelmeyer\inst{1} \and S.~Dreizler\inst{1} \and
  D.~Homeier\inst{1} \and A.~Reiners\inst{1,2}}

\institute{Institut f\"ur Astrophysik, Georg-August-Universit\"at
  G\"ottingen, Friedrich-Hund-Platz 1, 37077 G\"ottingen, Germany \and
  Hamburger Sternwarte, Universit\"at Hamburg, Gojenbergsweg 112,
  21029 Hamburg, Germany}

\date{Received $<$date$>$ / Accepted $<$date$>$}
 
\abstract
{Planets outside our solar system transiting their host star,
  i.~e. those with an orbital inclination near 90$^{\circ}$, are of special
  interest to derive physical properties of extrasolar planets. With
  the knowledge of the host star's physical parameters, the
  planetary radius can be determined. Combined with spectroscopic
  observations the mass and therefore the density can be derived from
  Doppler-measurements. Depending on the brightness of the host star,
  additional information, e.~g. about the spin-orbit alignment between
  the host star and planetary orbit, can be obtained.}
{The last few years have witnessed a growing success of transit
  surveys. Among other surveys, the MACHO project provided nine
  potential transiting planets, several of them with relatively bright
  parent stars. The photometric signature of a transit event is,
  however, insufficient to confirm the planetary nature of the faint
  companion. The aim of this paper therefore is a determination of the
  spectroscopic parameters of the host stars as well as a dynamical
  mass determination through Doppler-measurements.}
{We have obtained follow-up high-resolution spectra for five stars
  selected from the MACHO sample, which are consistent with transits
  of low-luminosity objects. Radial velocities have been determined by
  means of cross-correlation with model spectra. The MACHO light
  curves have been compared to simulations based on the physical
  parameters of the system derived from the radial velocities and
  spectral analyses.}
{We show that all transit light curves of the exoplanet candidates
  analysed in this work can be explained by eclipses of stellar
  objects, hence none of the five transiting objects is a planet.} 
{} 

\maketitle

\section{Introduction}

After the first detections of planets outside our solar system
\citep{1992Natur.355..145W,1995Natur.378..355M}, an intensive search
with various methods began \citep[see][for an
overview]{2002EuRv...10..185S} resulting in currently more than 200
planets (http://exoplanet.eu/). Most of these exoplanet detections
have been performed via the radial velocity (RV) method where the
``wobble'' of the parent star caused by the planet is measured by
spectral line shifts. Since these effects are very small for low-mass
planets in orbits of tens to hundreds of days, the determination of
orbital period, phase, inclination, eccentricity, and RV amplitude
demands RV accuracies of a few meters per second
\citep{2000ApJ...536L..43M}.

Meanwhile, alternative methods for planet detections have also been
successfully applied. The first four microlensing planets have been
detected
\citep{2004ApJ...606L.155B,2005ApJ...628L.109U,2006Natur.439..437B,2006ApJ...644L..37G},
possible first direct images of extra-solar planets were published
\citep{2004A&A...425L..29C,2005A&A...438L..25C,2005A&A...438L..29C,2005A&A...435L..13N,2006ApJ...641L.141B},
and the number of detections due to transit searches is steadily
increasing
\citep{2006ApJ...648.1228M,2006ApJ...acc..B,2006ApJ...651L..61O,2007MNRAS...acc..C}.

The transit method is of special interest, since it permits the
derivation of additional physical parameters of the planet, e.~g. the
radius can be measured either indirectly via the radius of the host
star or directly via detection of the secondary eclipse as observed
with the Spitzer Space Telescope
\citep{2005ApJ...626..523C,2005Natur.434..740D}. If combined with a
radial velocity variation measurement, the mass and mean density can
be determined, revealing constraints for the planetary
structure. Furthermore, transiting systems allow us to investigate the
atmospheres of the planets
\citep{2002ApJ...568..377C,2004ApJ...604L..69V} as well as the
spin-orbit-alignment between the rotational axis of the host star and
the planetary orbit
\citep{2006ApJ...acc..W,2006ApJ...acc..G,2006ApJ...653L..69W}.

\citet[hereafter DC]{2004ApJ...604..379D} published a list of nine
restrictively selected, transiting planet candidates from the MACHO
project \citep{1992ASPC...34..193A}. Only transit light curves with no
indication of gravitational distortion and only those with clear
U-shaped transit events were considered. De-reddened colours as well
as light curve fitting provide a good estimate of the companion
radius. Only companions below 3 Jupiter radii were selected.

Based on high-resolution spectra, the orbital velocities of five
potential host stars of exoplanet candidates have been measured. We
analysed the RV measurements together with MACHO transit light curves
in order to determine the system parameters complemented by a spectral
analysis.

The paper is organized as follows: In the next section, we shortly
describe the spectroscopic observations and the spectral analysis as
well as the Doppler-measurements. In section 3, we describe the
results of the individual systems and summarise in section 4.

\section{Observations and analyses}

In period 75 we secured three spectra for each of the five brightest
candidates. We used ESO's Fibre-fed Extended Range \'Echelle
Spectrograph (FEROS) mounted on the
\begin{table*}[ht!]
  \caption{Orbital elements, rotation velocities, and stellar
    parameters for all five analysed systems. Components {\it c} and {\it
      d} of MACHO 118.18272.189 and component {\it b} of MACHO
    120.22041.3265 are not visible in the spectra. $P_{\rm MACHO}$ is
    taken from \citet{2004ApJ...604..379D} and $P$ denotes the period
    derived using the light curves and RV measurements. $K$ is the
    semi-amplitude of the RV variations, $V_0$ the system velocity, and
    $i$ the orbital inclination. In case of systems with elliptical
    orbits, $e$ is the eccentricity and $\omega$ the longitude of the
    periastron. Furthermore, the mass $M^{\rm RV}$ is given in cases where
    the RV amplitude of two components is known. Then $T_{\rm eff}^{\rm
      RV}$ and $R$ are calculated for zero- and terminal-age main sequence
    models. $T_{\rm eff}^{\rm SA}$ is the effective temperature derived
    from the spectral analyses. In cases where just $T_{\rm eff}^{\rm SA}$
    from the spectral analyses is known, $M^{\rm SA}$ and $R$ are derived
    masses and radii from evolution models. All values in this table
    relate to the assumption of zero-age main sequence stars.}
  \label{table:1}
  \centering
  \begin{tabular}{l@{\,}l c c r@{\,$\pm$\,}l r@{\,$\pm$\,}l c c c c c c c c}
    \hline\hline\noalign{\smallskip}
    MACHO ID & & $P$ & $P_{\rm MACHO}$ & \multicolumn{2}{c}{$K$}
    & \multicolumn{2}{c}{$V_0$} & $i$ & $e$ & $\omega$ & $M^{\rm RV}$ &
    $T_{\rm eff}^{\rm RV}$ & $T_{\rm eff}^{\rm SA}$ & $M^{\rm SA}$ & $R$ \\
    & & [days] & [days] & \multicolumn{2}{c}{[km s$^{-1}$]} &
    \multicolumn{2}{c}{[km s$^{-1}$]} & [$^\circ$] & & [$^\circ$] &
    [$M_\odot$] & [K] & [K] & [$M_\odot$] & [$R_\odot$] \\
    \noalign{\smallskip}\hline\noalign{\smallskip}
    118.18272.189  & {\it a} & -- & 1.9673 &
    \multicolumn{2}{c}{0.00} & $-$25.51&0.03 & -- & -- & -- & -- & --  &
    5800 & -- & -- \\
    & {\it b} & -- & -- & \multicolumn{2}{c}{0.00} & $+$05.46&0.03 &
    -- & -- & -- & -- & -- & 5800 & -- & -- \\
    & {\it c} & 3.9346 & -- & \multicolumn{2}{c}{--} &
    \multicolumn{2}{c}{--} & (90.0) & -- & -- & 0.41 & 3730 & -- & --
    & 0.38 \\
    & {\it d} & 3.9346 & -- & \multicolumn{2}{c}{--} &
    \multicolumn{2}{c}{--} & & -- & -- & 0.41 & 3730 & -- & -- & 0.38
    \\[5pt]
    118.18407.57   & {\it a} & 4.7972 & 2.3986 & 78.84&0.10 &
    $-$20.48&0.07 & 84.0 & -- & -- & 1.27 & 6430 & 6200 & -- & 1.23 \\
    & {\it b} & -- & -- & \multicolumn{2}{c}{0.00} & $-$08.39&0.03 & &
    -- & -- & -- & -- & 6600 & -- & -- \\
    & {\it c} & 4.7972 & 2.3986 & 89.68&0.09 & $-$20.48&0.06 & & -- & --
    & 1.11 & 5980 & 6200 & -- & 1.04 \\[5pt]
    118.18793.469  & {\it a} & 4.0744 & 2.0372 & 75.81&0.18 &
    $-$56.30&0.11 & 85.6 & 0.041 & 89.94 & 0.90 & 5140 & 5400 & -- &
    0.81 \\
    & {\it b} & 4.0744 & 2.0372 & 83.67&0.25 & $-$56.30&0.14 & & & &
    0.82 & 5070 & 5400 & -- & 0.76 \\[5pt]
    120.22041.3265 & {\it a} & 5.4083 & 5.4083 & 22.18&0.06 &
    $-$24.00&0.04 & 89.8 & 0.108 & 19.98 & -- & -- & 6200 & 1.19 & 1.15 \\
    & {\it b} & 5.4083 & 5.4083 & \multicolumn{2}{c}{114.90}  &
    $-$24.00&0.04 & & & -- & -- & (3340) & (0.23) & 0.28 \\[5pt]
    402.47800.723  & {\it a} & 8.5496 & 4.2748 & 75.91&0.04 &
    $+$00.40&0.04 & 85.8 & -- & -- & 1.26 & 6400 & 6400 & -- & 1.22 \\
    & {\it b} & -- & -- & \multicolumn{2}{c}{0.00} & $-$26.40&0.04 & &
    -- & -- & -- & --& 5800 & -- & -- \\   
    & {\it c} & 8.5496 & 4.2748 & 68.09&0.07 & $+$00.40&0.07 & & -- & --
    & 1.40 & 6820 & 6400 & -- & 1.37 \\
    \hline
  \end{tabular}
\end{table*}
$2.2$~m telescope at La Silla, Chile. The spectrograph provides a
spectral resolution of ${\rm R} \sim 48\,000$ and covers a wavelength
range from $3500$~\AA\ to $9200$~\AA. The instrumental specifications
list a RMS velocity error of $\sim 25~{\rm m~s^{-1}}$. This is
sufficient to detect faint low-mass star companions and distinguish
them from sub-stellar companions, which was the primary aim of the
observations. The secondary aim is to use the spectra for a spectral
analysis in order to derive the stellar parameters of the host stars.

The observations of the five targets have been performed between August
19 and September 16, 2005. For each object we have obtained three
spectra with exposure times between $2400$~s and $3500$~s, depending
on the brightness of the object. The signal-to-noise ratio is $\sim
10$.

The data were reduced using the FEROS Data Reduction System (DRS). The
\'echelle spectra were bias and flat field corrected and wavelength
calibrated. The latter calibration was additionally quality checked by
cross-correlating the observation with a sky line spectrum. The
spectra were then corrected by applying relative wavelength
shifts. Barycentric and Earth rotation velocity influences to the
wavelengths are accounted for automatically by the DRS.

For the determination of the radial velocities we used the extracted
FEROS spectra and synthetic spectra of main sequence model stars
calculated from LTE model atmospheres using \verb!PHOENIX!
\citep{1999ApJ...512..377H} version 14.2. Both spectra were normalised
and relative fluxes were interpolated on a logarithmic wavelength
scale. A cross-correlation (CC) between a model with $\Teff =
5600~{\rm K}$ and observation was performed between $5000$~\AA\ and
$7000$~\AA. The CC was implemented using a grid with 200 steps of
$\Delta \log{\left( \lambda/[{\rm \AA}] \right)}=2.2 \cdot 10^{-6}$ in
each direction. This method turned out to be robust against the use of
different model spectra. We could identify up to three spectroscopic
components in our data. Each of the peaks in the CC was then fitted by
a Gaussian and the position of the maximum of the fit gives the radial
velocity. The errors of the RV measurements were calculated from the
standard deviation of the Gaussian plus the accuracy limit of FEROS of
$25~{\rm m~s^{-1}}$. These RV errors are in the range between $50$ and
$350~{\rm m~s^{-1}}$.

The CC function was also used to determine the projected rotation
velocities of the stars. We therefore applied a solar spectrum as
template convolved with rotational profiles following
\citet{2005oasp.book.....G}. This method allows to derive stellar
radii in binaries, assuming a synchronised orbit. In this analysis,
the determined rotational velocity $v \, \sin i$ is of the order of
the uncertainty in most cases, which, due to the low signal-to-noise
ratio, is about $5$~km~s$^{-1}$. These derived radii are consistent
with the ones obtained from main sequence models (see
Table~\ref{table:1}). Additional constraints for the radii of the
binary components visible in the spectra can therefore not be derived.

In order to spectroscopically identify the components of the analysed
systems, we again used the \verb!PHOENIX! model grid which ranges from
$4000$~K to $6800$~K in $\Teff$ and from $-1.5$ to $0.0$ in relative
solar metallicity $[{\rm Fe/H}]$. It should be noted that this is not
sufficient for a detailed abundance determination, which was not the
aim of this work. The surface gravity is kept constant at $\Lg =
4.5$. Knowing the RV of the individual components of a system, the
models were gauss-folded to the resolution of the observation and
shifted to their position in the observed spectrum. Depending on the
number of spectral components, all possible combinations of model
spectra were tested for each observed spectrum. A $\chi^2$-test was
used to identify the best fitting models. Given the low
signal-to-noise ratio of the spectra, we estimate an uncertainty of
about $400~{\rm K}$ in effective temperature.

In cases where we know the RV amplitudes for two components (MACHO
118.1407.57, 118.18793.469, 402.47800.723), $M \sin i$ is known for
both components. Assuming $i = 90^\circ$ for the first iteration, we
determined radii and effective temperatures ($T_{\rm eff}^{\rm RV}$ in
Table~\ref{table:1}) from interpolation of the Geneva model tracks
\citep{1993A&AS...98..523S} assuming zero-age main sequence (ZAMS) or
terminal-age main sequence (TAMS) stars. We then applied the eclipsing
binary simulation software
\nf\footnote{http://www.hs.uni-hamburg.de/DE/Ins/Per/Wichmann/Nightfall.html}
with the derived stellar and orbital parameters from the previous step
as input and calculated a best model fit to the R-band light curve and
radial velocity measurements simultaneously. We used the third light
contribution and the inclination as free parameters and calculated
$\chi^2$-values for the light curve fits assuming ZAMS and TAMS
stars. In a second iteration, we repeated the fit with the now known
inclination (see Fig.~\ref{fig:chi2}). For these three systems the so
derived effective temperature can be compared to the one of the
spectral analyses ($T_{\rm eff}^{\rm SA}$ in
Table~\ref{table:1}). Deviations are within our estimated
uncertainties and show the overall consistency of our main-sequence
solution.

In the other two cases (MACHO 120.22041.3265 and MACHO 118.18272.189),
the effective temperature from the spectral analysis was used to
derive masses and radii of each components, again assuming ZAMS and
TAMS stars. In the light curve simulations for the MACHO R-band
photometry we varied the inclination and the radius $R_2$ of the
potential transiting object, assuming ZAMS and TAMS primary stars.

\section{Results}

\begin{figure}
  \includegraphics[clip,width=8.8cm]{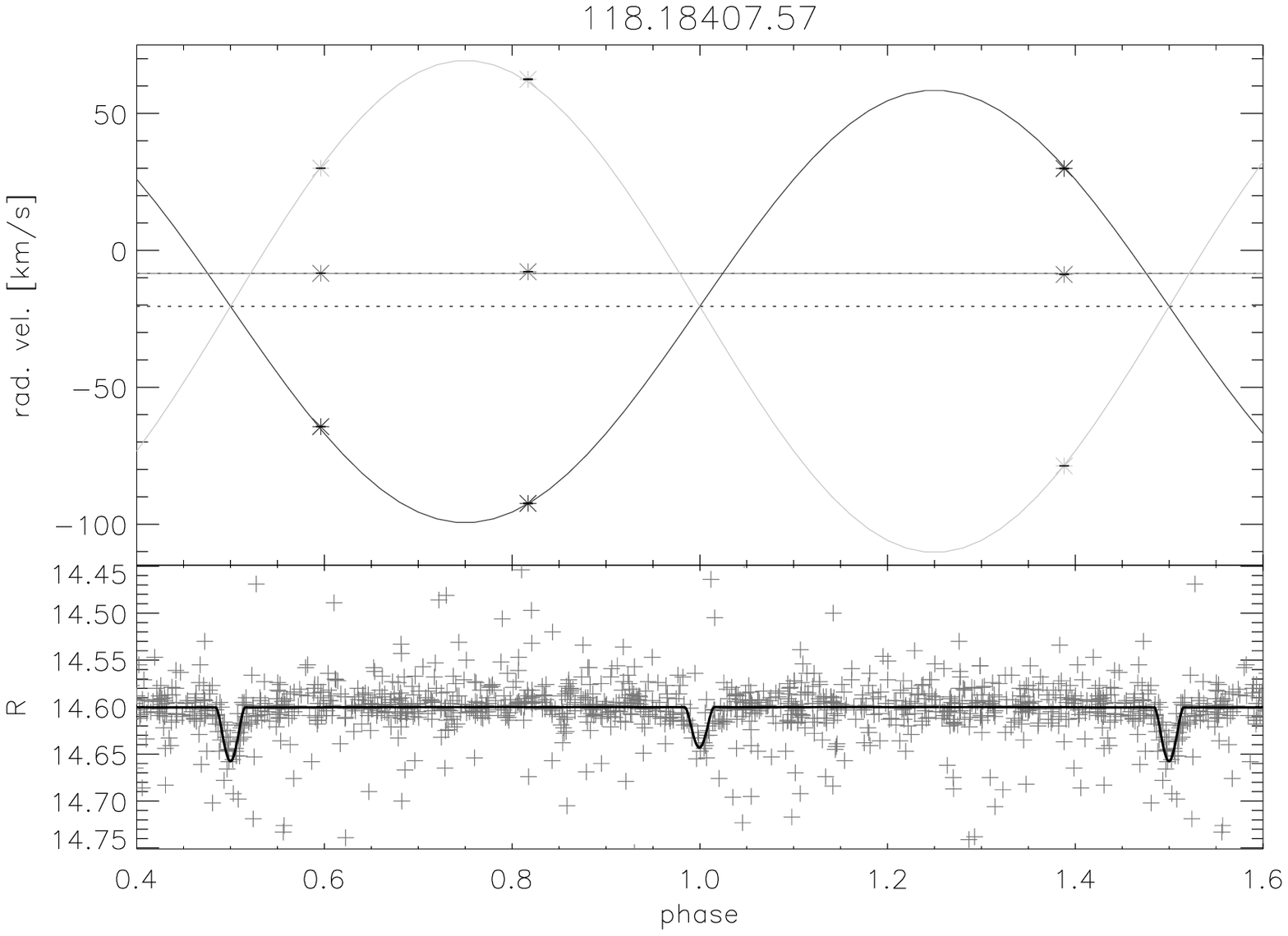}
  \includegraphics[clip,width=8.8cm]{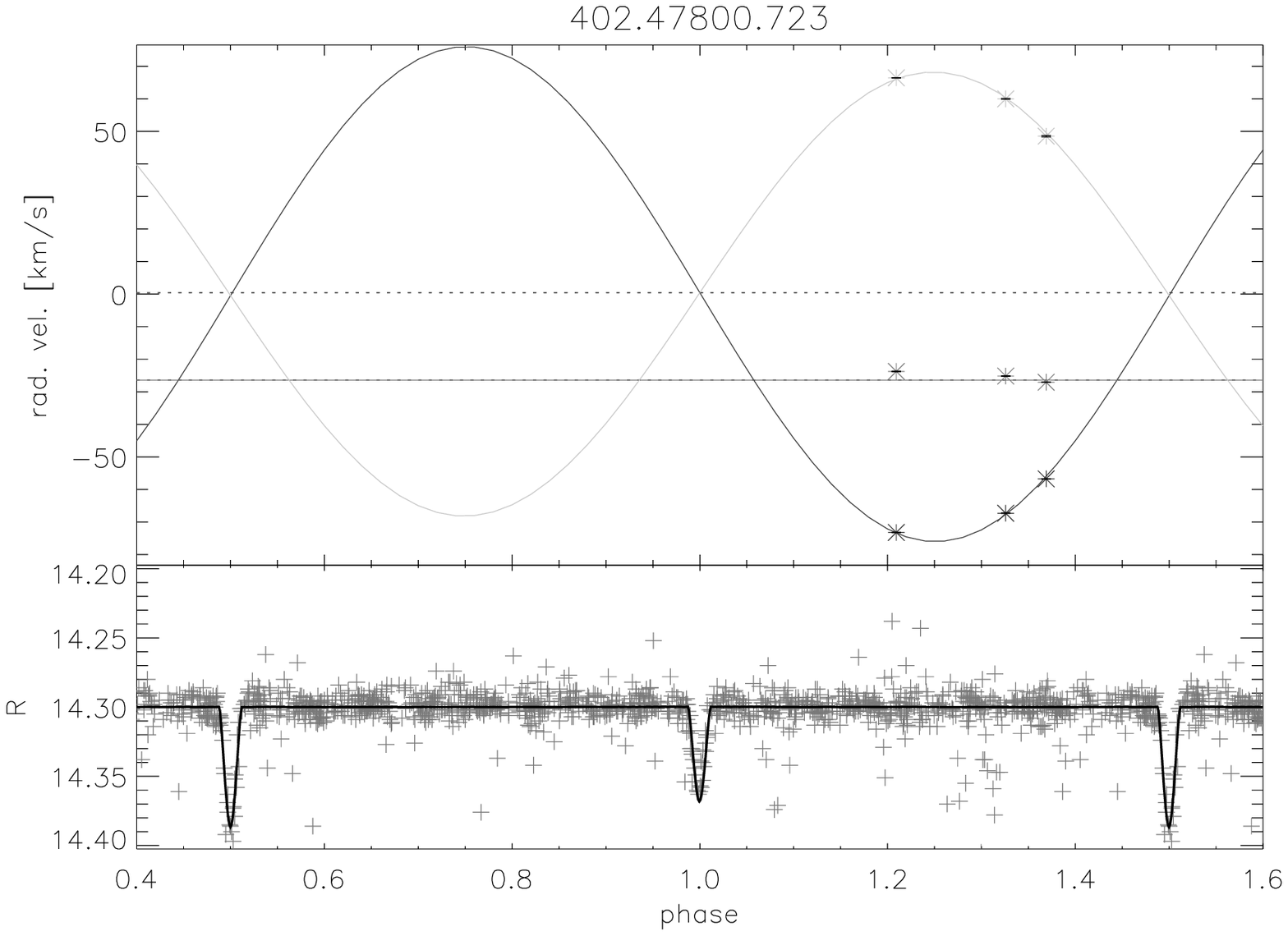}
  \includegraphics[clip,width=8.8cm]{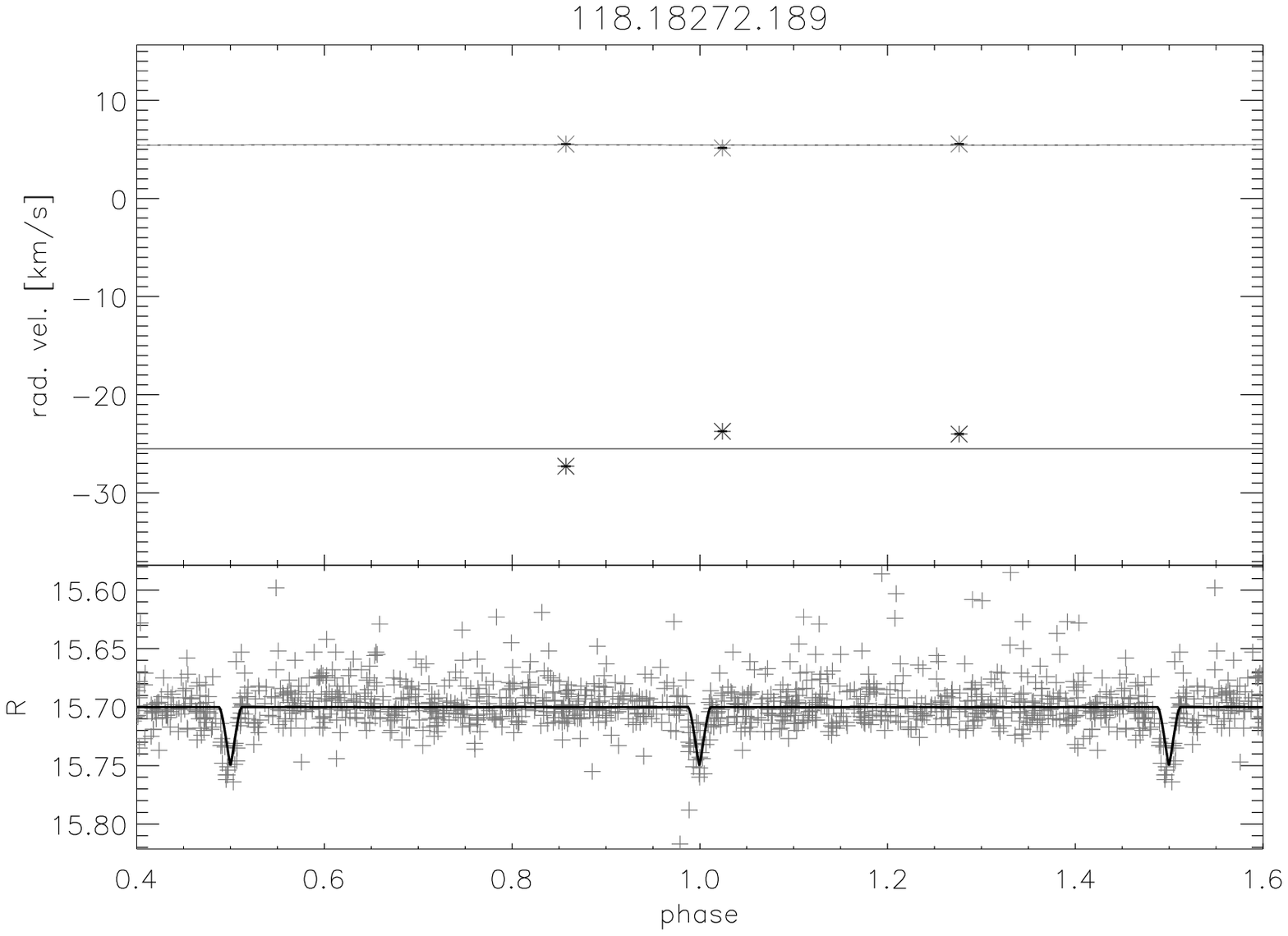}
  \caption{Radial velocity and light curve fits for systems with
    circular orbits. Component {\it a} is plotted in black, component
    {\it b} in grey, and component {\it c} in a lighter grey. The
    solutions shown are calculated assuming ZAMS stars. The error bars
    for the RV measurements are of the size of the symbols.}
  \label{fig:rvslc_circ}
\end{figure}

We will present results for the five targeted MACHO objects for which
we found an orbital solution that explains the detected transits and
the measured radial velocities. In Fig.~\ref{fig:rvslc_ell} we show
fitted light curves to the photometric MACHO data (bottom panels in
the plots) and RV curve fits to the Doppler-measurements (asterisks in
the top panel of the plots). The dashed lines are for circular orbits
and the solid lines show a best fit elliptical
orbit. Fig.~\ref{fig:rvslc_circ} again shows \nf\ light curve
solutions to the photometric data as well as the RV fits to the
Doppler-measurements. Here circular orbits reproduce the observations
best. All stars were assumed to be on the ZAMS for the fits in
Figs.~\ref{fig:rvslc_ell} and \ref{fig:rvslc_circ}. $\chi^2$-contour
plots for both ZAMS and TAMS stars are depicted in
Fig.~\ref{fig:chi2}. The inclination of the orbital plane and the
third light contribution (bottom three plots) and the radius $R_2$
(top two plots) of the potential transiting objects were treated to
vary. A list of the orbital parameters $P_{\rm MACHO}$ \citep[period
given by][]{2004ApJ...604..379D}, the derived period $P$ in our
analyses, the RV amplitude $K$, the system velocity $V_0$, the orbital
inclination $i$, and in cases of systems with elliptical orbits, $e$
the eccentricity and $\omega$ the longitude of the periastron as well
as the mass, effective temperature, and radius is shown in
Table~\ref{table:1}.

\subsection*{MACHO 120.22041.3265}

\begin{figure}
  \includegraphics[clip,width=8.8cm]{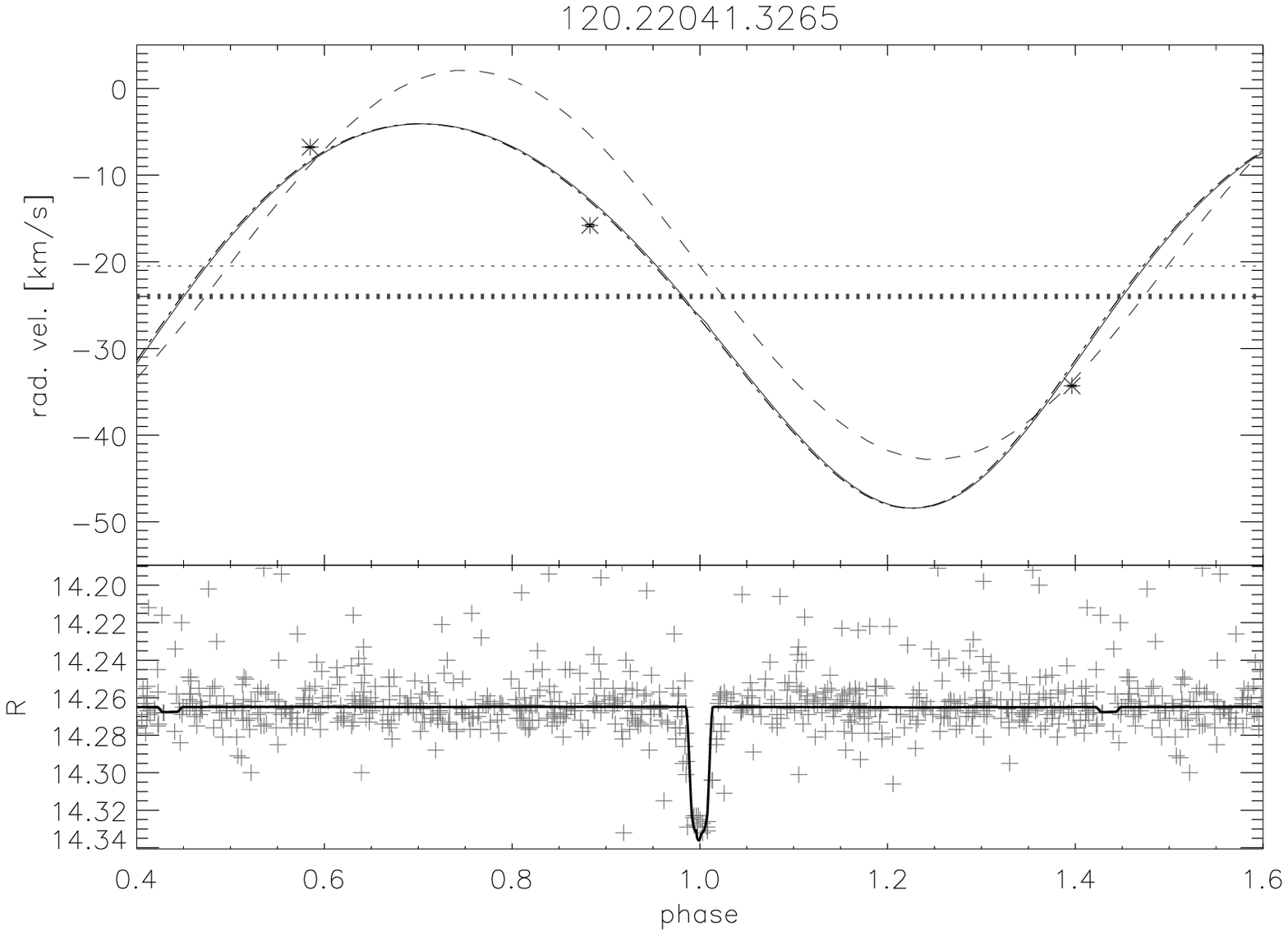}
  \includegraphics[clip,width=8.8cm]{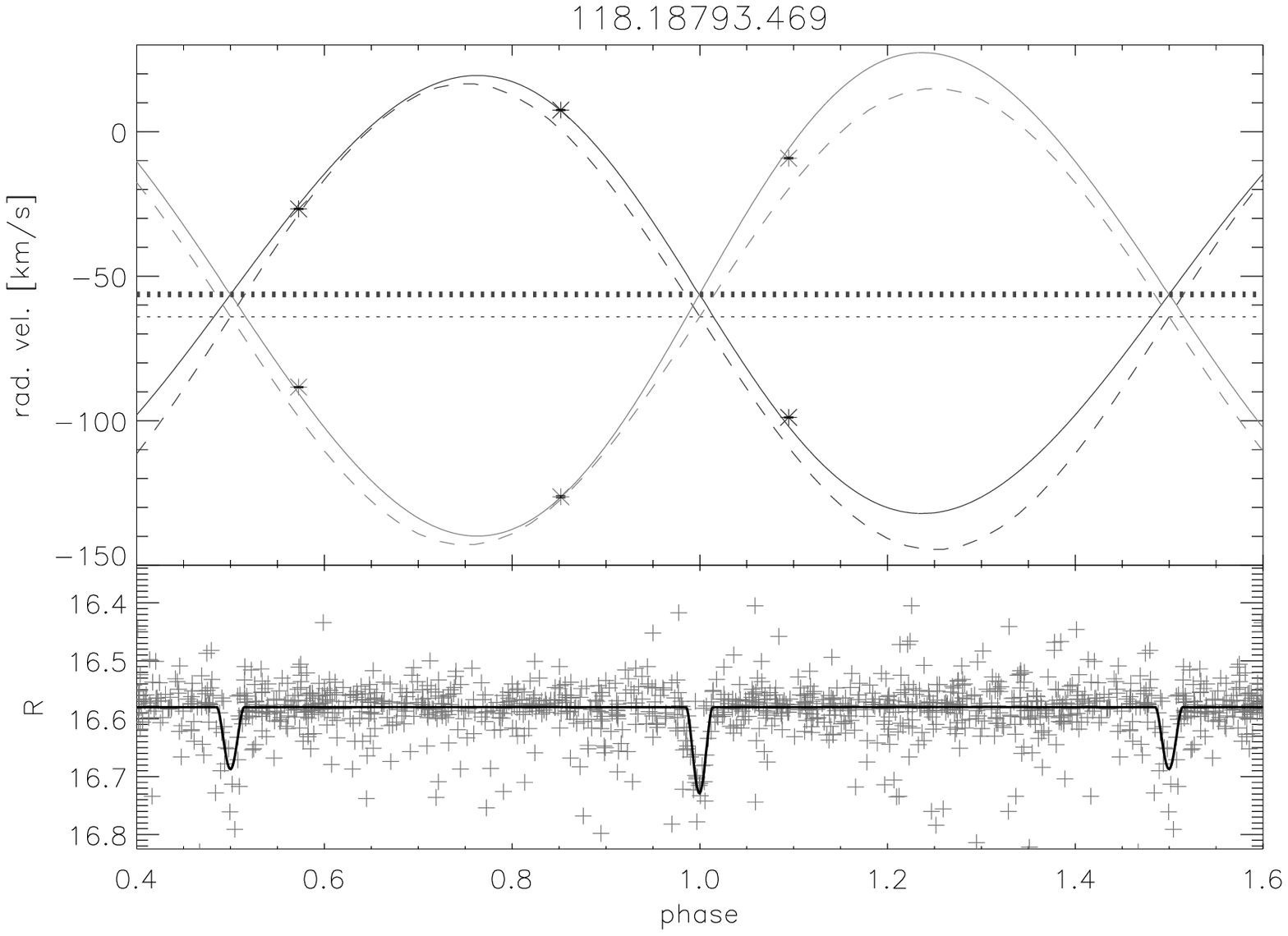}
  \caption{Radial velocity and light curve fits for systems with
    elliptical orbits. The dashed lines show best-fit sinusoidals while
    the solid lines show best-fit eccentric orbits. Component {\it a} is
    plotted in black, component {\it b} in grey. The system velocity for
    the circular orbit is shown by the thin line, and for the elliptical
    orbit by the thick dotted line. The solutions shown are calculated
    assuming ZAMS stars. The error bars for the RV measurements are of the
    size of the symbols.}
  \label{fig:rvslc_ell}
\end{figure}

MACHO 120.22041.3265 is the only system in our sample with just one
component visible in the spectra. Spectral analysis yields $\Teff =
6200~{\rm K}$ and indicates a low metallicity ($[{\rm
Fe/H}]=-1.0$). The fit of a sinusoidal to the RVs folded to a period
of $5.4083~{\rm d}$ (DC, dashed curve in Fig.~\ref{fig:rvslc_ell})
differs from the RV measurement at a phase of $0.87$ by $\sim 10~{\rm
km~s^{-1}}$. A better fit is provided by an elliptical orbit with an
eccentricity of $e=0.108$, a longitude of periastron of
$\omega=19.98$, and an orbital semi-amplitude of $K=22.18~{\rm
km~s^{-1}}$. For such a system the radius and mass of the secondary is
$R_2 = 0.3~\Rsol$ and $M_2 = 0.23~\Msol$ for a ZAMS and $R_2 =
0.5~\Rsol$ and $M_2 = 0.25~\Msol$ for a TAMS primary (see
Fig.~\ref{fig:chi2}), clearly indicating an M dwarf companion.
With these parameters, the system is very similar to OGLE-TR-78
\citep{2005A&A...438.1123P}.

We used equation~(6.2) of \citet{1977A&A....57..383Z} to calculate an
estimate for the circularisation time of the system. Due to the low
mass ratio $q=M_2/M_1$, we find a circularisation time of the order of
the Hubble time even for this close binary system.

\subsection*{MACHO 118.18793.469}

Two spectral components could be identified, each with
$\Teff=5400~{\rm K}$ and a highly subsolar ($[{\rm Fe/H}]=-1.5$)
metallicity. For the light curve and RV fits with \nf\ we used the RV
amplitudes of the two stars to derive masses by the procedure
described in the previous section. A reasonable fit to the RV
measurements folded to twice the period of DC can be achieved with
sinusoidals (dashed curves in Fig.~\ref{fig:rvslc_ell}),
i.~e. assuming a circular orbit for the two components. However, an
improved fit can be achieved by fitting the light curve and radial
velocities in \nf\ at the same time to an elliptical orbit (solid
curves in Fig.~\ref{fig:rvslc_ell}). The best fit is achieved with a
small eccentricity of $e = 0.041$ and a periastron longitude of
$\omega=89.94^\circ$. By varying the third light and the inclination,
we construct the $\chi^2$-map shown in Fig.~\ref{fig:chi2}. As
suggested by the spectral analysis of MACHO 118.18793.469, the lowest
$\chi^2$-value is found for a third light of zero. The inclination is
$85.6^\circ$ for the ZAMS and $82.2^\circ$ for the TAMS. The low
depths of the transit is therefore due to a grazing eclipse. This is
also supported by the V-shape of the best-fit model.

The best-fit light curve model shows different transit depths. This is
an indicator for two transits in one orbital period caused by two
stars of slightly different size.

\subsection*{MACHO 118.18407.57}

Three components are visible in the CCs of the three spectra, one of
which shows RV variations below $1~{\rm km~s^{-1}}$. Therefore, this
component is a third component, either in a wider orbit or physically
unrelated to the other two. Component {\it a} and {\it c} show RV
changes of over $100~{\rm km~s^{-1}}$. They can be well fitted with
sinusoidals of twice the period given by DC, i.~e. $4.7972~{\rm
d}$. If the photometric data are phased accordingly, we then see both
transits where the transit depths are reduced due to third light of
component {\it b}.

For the light curve simulation we once more used the RV amplitudes of
{\it a} and {\it c} to get the masses and varied the inclination and
third light coming from component {\it b}. The effective temperatures
and radii of the components are interpolated from the Geneva evolution
tracks assuming young stars on the ZAMS and older stars on the
TAMS. The contribution of component {\it b} meets the expectations
from the spectral analyses ($\Teff = 6200~{\rm K}$ for components {\it
a} and {\it c} and $\Teff = 6600~{\rm K}$ for component {\it b}, also
see Fig.~\ref{fig:chi2}). The inclination of the system is $84^\circ$
assuming that the stars are on the ZAMS and $79.5^\circ$ for the
TAMS. The system shows different transit depths, as MACHO
118.18793.469 does.

\subsection*{MACHO 402.47800.723}

The second brightest object in the sample shows three components in
the spectra. Components {\it a} and {\it c} are best fitted by a
model with $\Teff=6400~{\rm K}$, {\it b} has $\Teff=5800~{\rm K}$. As
in the case of MACHO 118.18407.57, the masses are derived from the
radial velocities. The RV measurements of {\it a} and {\it c} are well
fitted assuming a circular orbit with twice the period of DC. The
third component only shows small RV variations and therefore seems to
have a larger period than the other two. The fractional third light
contribution for this component is $\sim 1/3$ and an inclination of
the eclipsing system is $85.8^\circ$ assuming stars on the ZAMS and
$82.3^\circ$ for the TAMS (see Fig.~\ref{fig:chi2}). We again see
transit depth differences. Due to the high signal-to-noise ratio of
the light curve, these are quite obvious and amount to $\Delta R =
0.015$. This observation is also expected from the orbital
semi-amplitude differences.

\subsection*{MACHO 118.18272.189}

Each of the three FEROS spectra displays two components. The spectral
analysis reveals that both components have a similar effective
temperature ($\Teff = 5800~{\rm K}$) and a subsolar metallicity of
$[{\rm Fe/H}]=-0.5$. The cross-correlation shows that component {\it
b} has a constant RV of $\sim 5.46~{\rm km~s^{-1}}$ within the above
mentioned statistical errors. Component {\it a} shows RV variations of
$\sim 3.5~{\rm km~s^{-1}}$. Folding the RV measurements to the orbital
period given by DC, one sees that the two components visible in the
spectrum cannot be responsible for the transit in the light curve
since one RV point is very close to the transit.  However, here the
two components should almost have the same RV. This is clearly not
present in the data. The same is the case if we double the period (see
Fig.~\ref{fig:rvslc_circ}). Thus, we exclude the scenario that the two
visible components are responsible for the transit.

\begin{figure*}[ht!]
  \includegraphics[clip,width=18cm]{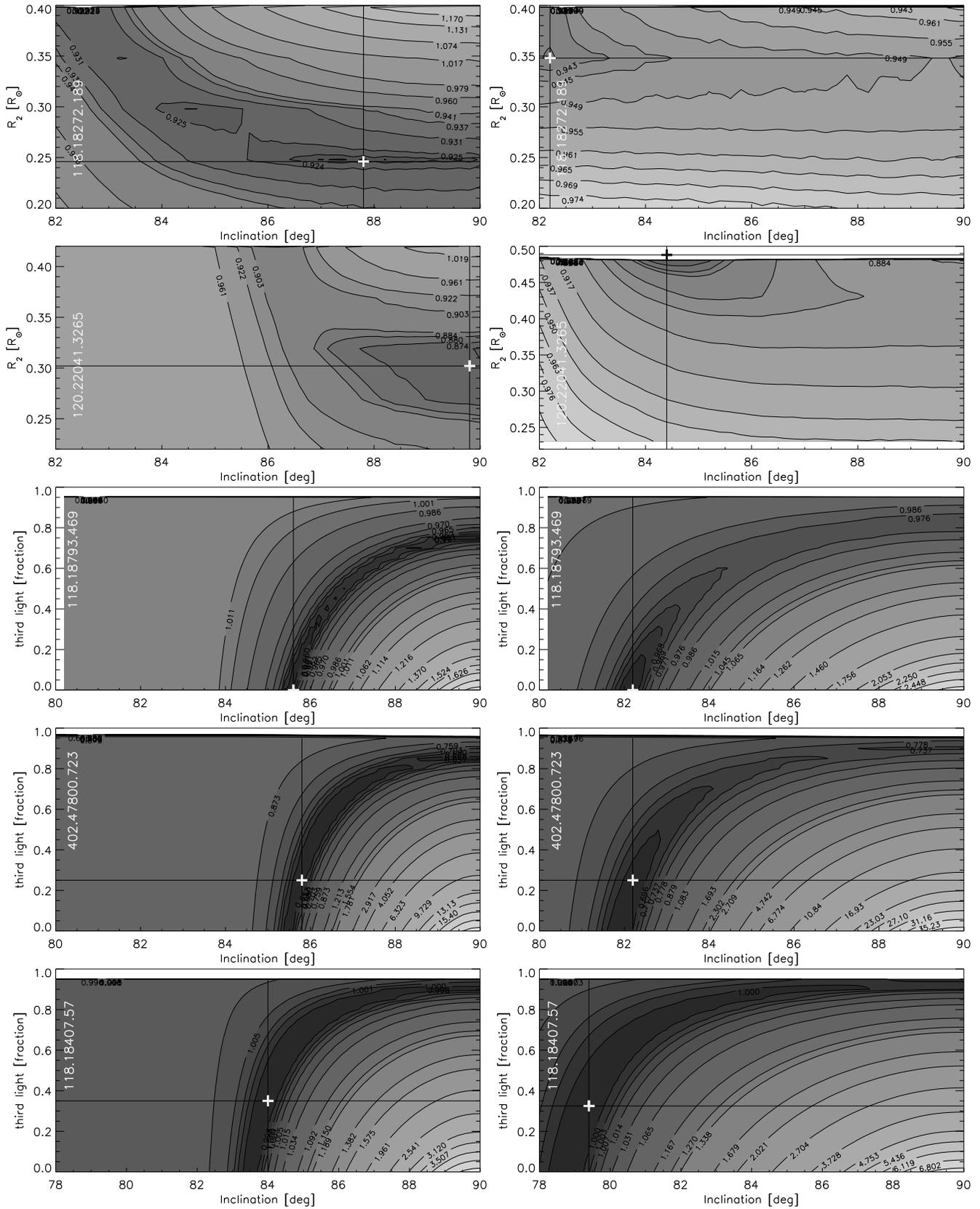}
  \caption{$\chi^2$-contour plots for all analysed systems. In the
  left column we assume that the stars are on the zero-age main
  sequence and in the right column on the terminal-age main
  sequence. The bottom three plots show the $\chi^2$-contours for
  third light and inclination as fitted parameters. For the top two
  the radius of the eclipsing component and the inclination have been
  varied. The crosses mark best-fit values.}
  \label{fig:chi2}
\end{figure*}

In another plausible scenario, we treat component {\it b} as third
light and assume that component {\it a} is eclipsed by a low mass
object not visible in the spectra. However, if we fit a sinusoidal to
the RV points, in our solution the star would move away from the
observer after the transit while it should do the opposite. We can
therefore discard this scenario.

One could argue that the variations in RV measured for component {\it
a} is just caused by systematic errors and that the eclipse visible in
the light curve is caused by a planet orbiting {\it a} without causing
any noticeable RV changes. We have fitted this scenario taking the
light from component {\it b} into account and found a radius of $R_2 =
0.25~R_\odot$ assuming that {\it a} is on the ZAMS and $R_2 =
0.35~R_\odot$ for {\it a} being on the TAMS (see
Fig.~\ref{fig:chi2}). These values, however, seem unrealisticly high
for planets and we can reject the 3-body scenario.

Finally, one scenario that can explain both the transit light curve
and the measured RVs is a four body system consisting of the two G
stars which are visible in the spectra and two M dwarfs invisible in
the spectra. Here the two faint components orbit each other in twice
the period from DC and eclipse each other twice. We assume an
inclination of $90^\circ$ and two low-mass stars of equal size. The
effective temperature of the eclipsing bodies was derived from the
transit depth of the MACHO R-band light curve using blackbody fluxes
for all four components. The transit depth is reduced by the light of
components {\it a} and {\it b}.  The RV variations of component {\it
a} can in this scenario be explained by the reflex motion of {\it a}
to the orbit of the binary M star system with a much larger period. We
therefore do not observe a correlation between the transits and the
RV. This scenario is underlined by the fact that the two RV
measurements in Fig.~\ref{fig:rvslc_circ} at periods of $\sim 1.0$ and
$\sim 1.3$, which have approximately the same RVs, are from two
spectra only taken one day apart, while the third RV value comes from
a spectrum 26 days later. Component {\it b} would be in a very wide
orbit or physically unrelated to the other three stars.

\section{Summary}

For none of the five analysed MACHO-candidates a planetary or brown
dwarf companion could be identified. We therefore confirm the
speculation of DC that due to the depths of the transits in the
photometric data the objects would be low-mass stars rather than
sub-stellar objects. From the five candidates, we found one grazing
eclipse of two nearly identical G stars (MACHO 118.18793.469), two
blends of deep eclipses of G stars with a significant third light
contribution (MACHO 118.18407.57 and MACHO 402.47800.723), one binary
star with a G type primary and an M dwarf secondary (MACHO
120.22041.3265) and one rather complicated, blended system with four
stars, of which each two are nearly identical (G and M type). With
this work we could show that also for deep transit surveys for
extrasolar planets, follow-up observations to weed out false positives
are efficiently possible with moderate effort.

After all, our results once again underline the need for spectroscopic
follow-up of transit planet candidates as already shown by
\citet{2005A&A...431.1105B} and \citet{2005A&A...438.1123P} for the
OGLE survey and \citet{2004ApJ...614..979T} in the case of a blend
scenario.

\begin{acknowledgements}
We would like to thank the referee for very useful comments.
\newline S.D.H. gratefully acknowledges the support of the
German-Israeli Foundation for Scientific Research and Development
grant I-788-108.7/2003.  
\newline A.R. has received research funding from the European
Commission's Sixth Framework Programme as an Outgoing International
Fellow (MOIF-CT-2004-002544).
\newline This paper utilizes public domain data obtained by the MACHO
Project, jointly funded by the US Department of Energy through the
University of California, Lawrence Livermore National Laboratory under
contract No. W-7405-Eng-48, by the National Science Foundation through
the Center for Particle Astrophysics of the University of California
under cooperative agreement AST-8809616, and by the Mount Stromlo and
Siding Spring Observatory, part of the Australian National University.
\end{acknowledgements}

\bibliographystyle{bibtex/aa}
\bibliography{macho}

\end{document}